\documentclass[aps,pre,superscriptaddress,showpacs,twocolumn]{revtex4-1}
\usepackage{epsfig}
\usepackage{color}
\usepackage{amssymb}
\usepackage{amsmath}
\usepackage[ngerman,english]{babel}
\usepackage{multirow}
\usepackage{graphicx}
\usepackage{subfigure}
\usepackage{float}
\usepackage{mathtools}
\usepackage{feynmf}
\usepackage{diagbox}

\newcommand{\etal}{{\it et al.}}
\newcommand{\pr}[4]{Phys. Rev. #1 {\bf #2}, #3 (#4)}
\newcommand{\hedp}[3]{High Energy Density Phys. {\bf #1}, #2 (#3)}
\newcommand{\astropj}[3]{Astrophys. J. {\bf #1}, #2 (#3)}

\newcommand{\physfluid}[3]{Phys. Fluids {\bf #1}, #2 (#3)}

\newcommand{\rmnumb}[2]{{#1}_{\rm #2}}
\newcommand{\rmupper}[3]{{#1}^{{\rm #2}}_{#3}}
\newcommand{\rmboth}[3]{{#1}^{{\rm #2}}_{\rm #3}}

\newcommand{\bks}[1]{\left( #1 \right)}

\newcommand{\squarebra}[1]{\left[ #1 \right]}

\newcommand{\biggcurlybra}[1]{\bigg\{ #1 \bigg\}}

\newcommand{\impart}[0]{{\rm Im}\, }
\newcommand{\repart}[0]{{\rm Re}\, }

\newcommand{\absvalue}[1]{\left| {#1} \right| }

\begin{document}

\title{Ionization potential depression and dynamical structure factor in dense plasmas}

\author{Chengliang Lin}
\affiliation{Universit\"at Rostock, Institut f\"ur Physik, 18051 Rostock, Germany}
\author{ 
Gerd R\"opke}
\affiliation{Universit\"at Rostock, Institut f\"ur Physik, 18051 Rostock, Germany}
\author{Wolf-Dietrich Kraeft}
\affiliation{Universit\"at Rostock, Institut f\"ur Physik, 18051 Rostock, Germany}
\author{Heidi Reinholz}
\affiliation{Universit\"at Rostock, Institut f\"ur Physik, 18051 Rostock, Germany}
\affiliation{University of Western Australia School of Physics, WA 6009 Crawley, Australia}
\date{\today}

\begin{abstract}
The properties of a bound electron system immersed in a plasma environment are strongly modified 
by the surrounding plasma. The modification of an essential quantity, the ionization energy, 
is described by the electronic 
and ionic self-energies including dynamical screening within the framework of the quantum statistical theory.
Introducing the ionic dynamical structure factor as the indicator for the ionic microfield,  
we demonstrate that  ionic correlations and fluctuations  play a critical role in determining the 
ionization potential depression. This is in particular true for mixtures of different ions 
with large mass and charge asymmetry. 
The ionization potential depression is calculated for dense aluminum plasmas as well as for a CH plasma 
and compared  to the experimental data and more phenomenological approaches used so far.
\end{abstract}

\maketitle

%\noindent Keywords:
%\begin{keyword}
%non-equilibrium statistical operator, reduced statistical operator,
%quantum master equation, quantum Brownian motion, open quantum
%systems, decoherence, Rydberg atoms, ultracold plasma \\
% PACS number(s): 03.65.Yz,52.25.Gj,52.25.Mq
%\end{keyword}

%%%%%%%%%%%%%%%%%%%%%%%%%%%%%%%%%
%%%%%%%%%%%%%%%%%%%%%%%%%%%%%%%
%%%%%%%%%%%%%%%%%%%%%%%%%%%%%%%%%%
\section{Introduction}
In the context of new experimental facilities exploring warm dense matter (WDM) 
and materials in the high-energy density  regime, 
a detailed theoretical investigation of thermodynamic, transport and optical properties 
of strongly coupled and nearly degenerate Coulomb systems becomes of emerging interest. 
This is of relevance not only for material science investigating matter 
under extreme conditions (Mbar pressures, temperatures of 1 eV up to 1 keV),
like inertial confinement fusion implosions in laboratory experiments, 
but also for understanding the structure and evolution of the increasing number 
of known planets as well of other astrophysical objects. 
%Optical spectra modified by a plasma environment 
%are of relevance in laboratory physics as well as in astrophysics. 

A fundamental phenomenon is the modification of bound state levels 
as well as of continuum states owing to the surrounding warm and dense medium. 
Here, we are interested in the ionization potential depression (IPD) 
which is relevant for the composition of the plasma, and, in this way, 
for the thermodynamic and transport properties.
We focus on experiments showing the dissolution of spectral lines due to the IPD 
which determines the ionization degree of WDM.
Accurate predictions are necessary for simulation codes such as FLYCHK \cite{Chung05} 
which model plasmas under extreme conditions.

Being a long-standing problem in plasma physics, 
IPD %new 
experiments \cite{Hoarty13,ciricosta12,ciricosta16,preston13,Fletcher14,Kraus16} 
have been performed recently using the new possibility to produce 
highly excited plasmas at condensed matter densities 
by intense short-pulse laser irradiation.
Comparisons of observed optical spectra with simulations using traditional expressions 
for the IPD given by Ecker and Kr\"oll (EK) \cite{EK63}
or Stewart and Pyatt (SP) \cite{SP66} have been performed. 
Neither of them leads to a satisfying description for all of the available experiments. 
While, on one hand, Hoarty's results \cite{Hoarty13} on the disappearance of spectral lines seem to favor SP, 
and, on the other hand, the direct measurements on the ionization energy of the K-shell in aluminum and the subsquent 
K$_\alpha$ lines by Ciricosta \textit{et al.} \cite{ciricosta12,ciricosta16} tend to confirm EK, 
recently reported results by Kraus \etal~\cite{Kraus16} can not be understood by either of the two approaches. 
A more systematic and accurate theory is demanded to describe the measurements.

The commonly used expressions for the IPD derived by Ecker and Kr\"oll \cite{EK63} or
Stewart and Pyatt \cite{SP66} interpolate 
between the Debye (DH) limit for low  densities and an ion sphere (IS) expression, see \cite{ZM80},
for high densities. They are based on simplified assumptions such as the introduction 
of an average static potential to perform Thomas-Fermi calculations. 
A critical discussion of these approaches and their applicability for the experiments 
given above was presented in \cite{Crowley14}. 
Other approaches use Hartree-Fock-Slater calculations \cite{STJZS14}, Monte Carlo simulations \cite{Stransky16}, 
molecular dynamics simulations \cite{Calisti15}, density-functional theory calculations \cite{VCW14},
microfield concepts and a detailed configuration accounting description \cite{IS13,IS14}, or the theory 
of disordered solids where itinerant band electrons become localized below a mobility edge \cite{DP92}.

A systematic approach to describe the properties of dense plasmas is given 
by the quantum statistical many-body theory, 
in particular the use of the Green function method \cite{KKER}. 
It has been applied to optical properties \cite{GHR91} by calculating shifts and broadening of  
spectral lines in a plasma environment. The shift of bound states and the 
continuum edge in dense plasmas has also been considered in Refs. \cite{Seidel,RKKKZ78,ZKKKR78}.

Already some decades ago, the shifts both of the continuum edge and of the bound state levels have been discussed  for
the electron-hole plasma in excited semiconductors \cite{RKKKZ78,ZKKKR78,KKER,Zimmermann,KSK}.
Depending on the density and temperature of the electron-hole plasma, excitons are 
modified by medium effects, and merge with the lowered continuum at the Mott density.
Thus, an exciton gas is transformed into an electron-hole liquid. A highly sophisticated 
theory describing dynamical screening and degeneracy effects 
by the fermionic plasma constituents had been worked out, explaining precise
measurements in excited semiconductors. 
However, because the ions are heavier compared 
to the effective mass of holes, a simple transfer of the physics of excited semiconductors 
to WDM is not possible. The  ions remain classical within a large density region, forming strong correlations  
which are described by the dynamical ionic structure factor (SF)
$S_{\rm ii}({\bf q},\omega)$.

In the following, we will give a relation between the IPD and the ionic structure factor. Thus,
mean-field (average atom) approaches are improved taking into account fluctuations of the ionic microfield.
Further systematic improvements would be possible considering higher order Feynman diagrams in the Green function approach.

%%%%%%%%%%%%%%%%%%%%%%%%%%%%%%%%%%%%%%%%
%%%%%%%%%%%%%%%%%%%%%%%%%%%%%%%%%%%%%%%%
%%%%%%%%%%%%%%%%%%%%%%%%%%%%%%%%%%%%%%%%
\section{The in-medium two-particle problem}
\label{sec:2}
We consider a two-particle system, consisting of an electron (charge $-e$, mass $m_e$)
and an ion (charge $\bks{Z_i +1}e$, mass $m_i$) imbedded in a surrounding
plasma. In vacuum, the solution of the Schr\"odinger equation for the Coulomb interaction is well known. 
Bound states are found at negative energies, whereas a continuum of scattering states is observed at positive energies.
The simple case of the hydrogen atom can be generalized to a two-particle system 
with total charge $Z_i e$, consisting of a core ion with charge number $Z_i+1$ and an electron, 
charge number $Z_e=-1$. 
According to these definitions, the notations $Z_i$ and $Z_i+1$ denote the charge number of the ions before 
and after the ionization, respectively. (Note, that the charge number $Z_i$ is 
at least by one smaller
than the nuclear charge number of the corresponding atom. For neutral atoms before ionization, we have obviously $Z_i =0$.)

If the two-particle system is embedded in a plasma, bound state energies and wave functions as well as 
the scattering states are modified. A systematic quantum statistical approach to describe these medium effects 
is given by the method of thermodynamic Green functions \cite{KKER,KSK}. 
In particular, the following in-medium Schr\"odinger equation (or Bethe-Salpeter 
equation) can be derived \cite{RKKKZ78,ZKKKR78,KKER,KSK}:
% 
% the Green function approach  leads to the following two-particle Bethe-Salpeter 
% equation \cite{RKKKZ78,ZKKKR78,KKER,KSK}
\begin{eqnarray}
\label{BSE}
&&\Big[E(1)+E(2)+ \sum_{\bf q}[f(1+{\bf q})+f(2-{\bf q})] V_{12}({\bf q})\nonumber\\&&
 + \Delta V^{\rm eff}(1,2,{\bf q},z) \Big ] \! \psi(1,2,z)
+\! \sum_{\bf q}\! \Big\{ [1\!-\!f(1)\!-\!f(2)] V_{12}({\bf q})  \nonumber\\&&
 + \Delta V^{\rm eff}(1,2,{\bf q},z)\Big\} \psi(1+{\bf q},2-{\bf q},z)=\hbar z \, \psi(1,2,z).
\end{eqnarray}
Here, the single particle states $1=\{ \hbar {\bf p}_1, \sigma_1,c_1\}$ are given by momentum, spin and species, 
respectively, $E(1) = \hbar^2 {\bf p}_1^2/(2 m_1)$. 
In the case considered here, $c_1$ and $c_2$ denote the electron and the core ion, respectively. 
For the interaction we assume the Coulomb potential $V_{12}({\bf q}) = Z_{c_1}Z_{c_2} e^2/(\varepsilon_0 q^2)$ 
which contains the charge numbers of the interacting particles, in our case $Z_{c_1}Z_{c_2}=-(Z_i +1)$.

The complex variable $z$ describes the analytical continuation of 
the functions, defined for the Matsubara frequencies, into the entire $z$ plane. 
Of interest is the behavior of the functions near the real axis, $z=\omega \pm i \epsilon$.

Neglecting in Eq. (\ref{BSE}) the medium effects arising from the effective interaction $\Delta V^{\rm eff}(1,2,{\bf q},z)$ 
as well as
the Fermi distribution functions $f(i)=[\exp(\beta (E(i)-\mu(i))+1]^{-1}$ 
with 
$\beta = 1/(k_B T)$ and $\mu(1)$ denoting the chemical potential of species 
$c_1$, the equation 
\begin{eqnarray}
\label{BSE0}
&&\Big[E(1)+E(2) \Big ] \! \psi(1,2,z)
+\! \sum_{\bf q}\! V_{12}({\bf q}) \psi(1+{\bf q},2-{\bf q},z)
\nonumber \\ &&=\hbar z \, \psi(1,2,z)
\end{eqnarray}
has eigensolutions $\psi_n(1,2)$ at energies $\hbar z = E_n$, well known from hydrogen-like ions. 
For more complex ions consisting of a nucleus and some bound electrons, 
a pseudopotential can be introduced to describe the effect of the electrons within the core ion.

The in-medium Schr\"odinger equation (\ref{BSE}) describes the influence of the medium by two effects, 
Pauli blocking and screening. Pauli blocking is caused by the antisymmetrization of the fermionic wave function.
States which already are occupied by the medium are blocked and can not be used 
for the two-particle system under consideration. The blocking is described by 
the Fermi distribution function. 
Pauli exclusion principle is acting as Fock shift $\sum_{\bf q}f(1+{\bf q})\, V_{12}({\bf q})$ in addition to the 
single particle energy $E(1)$ in Eq. (\ref{BSE}) (for charge-neutral plasmas, the Hartree term vanishes). Also in the interaction term, Pauli blocking gives the contribution 
$-\sum_{\bf q} f(1) V_{12}({\bf q}) \psi(1+{\bf q},2-{\bf q},z)$. Both in-medium 
contributions are caused by the degeneracy of the plasma particles. In the 
plasmas considered here, electrons may be degenerate 
because of their small mass $m_e$. The ions are non-degenerate and can be 
treated as classical particles.

Considering only the Pauli blocking effects, the effective (non-hermitean) Hamiltonian 
of Eq. (\ref{BSE}) remains real and can be symmetrized. The energy eigenvalue problem can be solved, 
and the bound state energies as well as the edge of continuum states are shifted. At a certain density, 
the bound states merge with the continuum of scattering states and disappear. 
Within this approximation, which is essentially a mean-field approximation, 
a sharp value for the lowering of the continuum edge and for the IPD can be calculated.

Screening of the interaction by the medium is  described by the effective interaction 
\begin{eqnarray}
\label{Veff}
&&\Delta V^{\rm eff}(1,2,{\bf q},z)= -V_{12}({\bf q})\int_{-\infty}^\infty \frac{d \omega'}{\pi}
{\rm Im}\, \varepsilon^{-1}(q,\omega'+i0)\nonumber \\&&
 \times\left[\rmnumb{n}{_B}(\omega')+1\right]
 \left( \frac{\hbar}{\hbar z-\hbar \omega'-E(1)-E(2-{\bf q})}\right.\nonumber \\&&
\left.+\frac{\hbar}{\hbar z-\hbar \omega'-E(1+{\bf q})-E(2)}
\right),
\end{eqnarray}
where terms $\propto f(1)$, which give corrections in higher orders of the density, are neglected. 
$\rmnumb{n}{_B}(\omega)=[\exp(\beta \hbar \omega)-1]^{-1}$ is the Bose distribution function. 
The dynamical properties of the surrounding plasma are contained in the dielectric function 
$\varepsilon(q,z)$ to be taken at the real axis, $z=\omega'+i0$. In general, this is a complex, 
frequency dependent quantity, with a jump of the imaginary part at the real axis. 
Often the random-phase approximation (RPA) is taken, and in the static limit $\omega \to 0$ the 
Debye screening is obtained. In this work, we show that these simple approximations have to be improved 
in a systematic way which is obtained from the quantum statistical approach.

Including the effective potential, the effective Hamiltonian in the in-medium Schr\"odinger equation (\ref{BSE})
becomes complex and frequency dependent. As a consequence, the eigenstates are no longer stationary
states with sharp energy levels which are shifted by the polarisation of the medium, but have a finite life time
given by the imaginary part of the effective Hamiltonian. 
This can be interpreted as collisions with the plasma particles and leads to a broadening of the energy levels. 
The corresponding quantum statistical approach to plasma line shapes based on the treatment of the 
polarization function has been worked out \cite{GHR91} and will not be investigated in the present work.

Subsequently, sharp level shifts and a sharp shift of the continuum edge are only obtained from a mean-field 
approximation. Any frequency dependence beyond the mean-field approximation gives imaginary parts and, 
in this way, a broadening of the continuum edge and the energy levels.
The latter problem has been considered also earlier \cite{ZKKKR78,Seidel} where both, real part and imaginary part
of the energy levels of the in-medium two-particle problem, are calculated. 
As a consequence, only the spectral function has a unique physical meaning, showing the spectral line profiles 
and the smooth transition to the continuum. However, within this work we will focus on the shifts 
that are obtained from the real part of the effective Hamiltonian.

As shown in \cite{KKER,KSK,RKKKZ78,ZKKKR78,Seidel}, density effects arise from dynamical screening 
in the effective potential, 
expressed by the inverse dielectric function $\varepsilon^{-1}(q,z)$ of 
the medium in Eq.\,(\ref{Veff}).
For bound states, Pauli blocking as well as the screening in the 
self-energy term ($\Delta V^{\rm eff}$ in the first square bracket of Eq.\,(\ref{BSE})) 
and the effective interaction partially compensate each other so that the bound state 
energy levels are only weakly dependent on the density.
In contrast, the energy shift of the continuum states is determined only by the self-energy contribution.
Therefore, in leading order of the density, the medium modification of the IPD is given by the shift of the edge of 
continuum states. For a more extended discussion see \cite{RKKKZ78,ZKKKR78,Zimmermann,Seidel,KKER,KSK}.

A standard expression for the dielectric function $\varepsilon(q,z)$ is the random phase approximation (RPA).
From the real part of the self-energy, the Debye shift of the continuum edge is immediately observed.
Here we discuss improvements beyond RPA to evaluate the shift of the continuum edge occurring at ${\bf p}_1={\bf p}_2=0$. 
Thus, our approach, which is based on a systematic quantum statistical approach, can be regarded as an improvement of
the Debye theory or other approaches using semi-empirical assumptions such as the ion sphere model.

%%%%%%%%%%%%%%%%%%%%%%%%%%%%%%%%%%%%%%%%%%%%%%%%%%%%%%%
%%%%%%%%%%%%%%%%%%%%%%%%%%%%%%%%%%%%%%%%%%%%%%%%%%%%%%%%%
\section{Shift of single particle states}
\subsection{Self-energy of single particle states}
In the single-particle picture, the influence of the plasma environment on the properties 
of the investigated particle is merged into the self-energy $\Sigma_c(1,z)$.
It can be represented by Feynman diagrams, in lowest approximation 
by the diagram (also known as $V^s G$ or $GW$ approximation) with the dressed  propagator $G$ and the screened
potential $V^s$
\begin{align}\label{GW}
 \Sigma_c(1,z) & = \sum_{{\mathbf{q}},\omega } G_c({\mathbf{p-q}},z-\omega)\cdot V^s({\mathbf{q}},\omega) \\ & \nonumber \\
 & =
 \begin{fmffile}{MW8799}
\begin{fmfgraph*}(15,-3)
\fmfleft{i1}
\fmfright{o1}
\fmf{fermion}{i1,o1}
\fmf{photon,right,tension=0}{o1,i1}
\end{fmfgraph*}
\end{fmffile}
\, = \rmupper{\Sigma}{HF}{c}(1,z) + \rmupper{\Sigma}{corr}{c}(1,z)\nonumber .
\end{align}
The Hartree-Fock (HF) contribution to the self-energy has been investigated elsewhere, see \cite{KKER}, and will not be 
discussed here.
The correlation part of the self-energy $\Sigma^{\rm corr}_c(1,z)$ 
contains the contribution of the interaction with electrons, as well as the interaction with ions. 
We are interested in the real part of the self-energy since it describes the continuum shift.  
It follows from Eq.\,(\ref{Veff}) by renaming, e.g., $\hbar z -E(2)=\hbar \omega$ in the last term of (\ref{Veff}). Then we have
\begin{eqnarray} \label{deltasigmaeps}
&& \repart \rmupper{\Sigma}{corr}{c}(p,\omega) = - {\cal P}\, \int \frac{d^3 {\bf q}}{(2\pi)^3}
 \int \frac{ d \omega'}{\pi} V_{cc}(q) \nonumber \\ 
 &&\times \impart \varepsilon^{-1}(q,\omega'+i0)
 \frac{1+\rmnumb{n}{_B}(\omega')}{ \omega-\omega'-E_{c,{\bf p}+{\bf q}}/\hbar}.
\end{eqnarray}
($ {\cal P}$ denotes the principal value.)
In general, the dielectric function is connected to the %charge 
dynamical SF via the fluctuation-dissipation theorem. %For example, f
For a two-component plasma (free electrons with charge $-e$,  ions with effective charge $Z_i e$ and charge 
neutrality $Z_i n_i=n_e$),
the imaginary part of the inverse dielectric function can be expressed via the dynamical SFs, see also \cite{Chihara},
\begin{eqnarray}
\label{Imeps}
&& \impart \varepsilon^{-1}({\bf q},\omega+i0) 
 = \frac{e^2}{\varepsilon_0\, q^2} \frac{\pi}{\hbar\, \bks{1+\rmnumb{n}{_B}(\omega)}}
  \\&&
\times \left[ Z_i^2\,n_i S_{\rm ii}({\bf q},\omega) 
 -2Z_i \,\sqrt{n_e n_i}S_{\rm ei}({\bf q},\omega)+n_e S_{\rm ee}({\bf q},\omega) \right]\nonumber\,.
\end{eqnarray}
The dynamical SFs $S_{cd}({\bf q},\omega)$ characterize the plasma in response to any perturbation. 
For instance, they have been investigated to describe
X-ray Thomson scattering, see Ref. \cite{GRHGR07}. Other plasma properties such as the electrical conductivity 
are also governed by the dynamical SF. 
The dynamical SFs are related to the density-density correlation functions
$\langle \delta n_c({\bf r},t) \delta n_d(0,0) \rangle$ via Fourier transformation. Note, that it is also connected to
the symmetrized correlation function of the longitudinal microfield fluctuations 
$\langle \delta {\bf E} \delta {\bf E} \rangle_{{\bf q},\omega}$ \cite{KSK}
\begin{equation}
 \langle \delta {\bf E} \delta {\bf E} \rangle_{{\bf q},\omega}=2 \pi (Z_i^2 e^2/q^2) S_{\rm ii}({\bf q},\omega).
\end{equation}

% obtained from the density-density correlation functions
%$\langle \delta n_c({\bf r},t) \delta n_d(0,0) \rangle$ after Fourier transformation. They are the fundamental
%quantities that describe the response of the plasma to any perturbation. For instance, they have been investigated to describe
%X-ray Thomson scattering, see Ref. \cite{GRHGR07}. Other plasma properties such as the electrical conductivity 
%are also governed by the dynamical SF. 
%
%Mention that the dynamical SF is related to the symmetrized correlation function of the longitudinal microfield fluctuations,
%$\langle \delta {\bf E} \delta {\bf E} \rangle_{{\bf q},\omega}=2 \pi (e^2/q^2) S_{\rm ii}({\bf q},\omega)$; see Ref. \cite{KSK}.
%
%%%%%%%%%%%%%%%%%%%%%%%%%%%%%%%%%%%%%%%%%%%%%%%%%%%%%%%%%%%%%%%%%
%%%%%%%%%%%%%%%%%%%%%%%%%%%%%%%%%%%%%%%%%%%%%%%%%%%%%%%%%%%%%%%%%%%%%%%%

For further discussion of the general expressions (\ref{deltasigmaeps}) and (\ref{Imeps}),
we perform exploratory calculations using model approaches for the dynamical SFs.
Following the relations for the dynamical SFs reported in Refs. \cite{Chihara,GRHGR07}
\begin{align}
S_{\rm ei}(q,\omega) & = \frac{\rmnumb{q}{sc}(k)}{\sqrt{Z_i}} S_{\rm ii}(q,\omega)  \\
S_{\rm ee}(q,\omega) & = S_{\rm ee}^0(q,\omega) + \frac{\absvalue{\rmnumb{q}{sc}(k)}^2}{Z_i} S_{\rm ii}(q,\omega), \nonumber
\end{align}
the decomposition of the dynamical SF as introduced in Eq. (\ref{Imeps}) can be  
divided into $S_{\rm ee}^0(q,\omega)$ of the fast moving free electrons 
and the ionic part $S_{\rm ii}^{ZZ}(q,\omega)$ which includes also the screening cloud of the slowly 
moving electrons following the ionic motion, denoted by $\rmnumb{q}{sc}(k)$, 
\begin{align}
 & Z_i S_{\rm ii} (q,\omega)
-2 \,\sqrt{Z_i}S_{\rm ei}(q,\omega) +S_{\rm ee}(q,\omega)  \nonumber \\
= & Z_i\,S_{\rm ii}^{ZZ}(q,\omega)+S_{\rm ee}^0(q,\omega)
\end{align}
with $S_{\rm ii}^{ZZ}(q,\omega) = \bks{1-\rmnumb{q}{sc}(k)/Z_i}^2 S_{\rm ii}(q,\omega)$.
The electronic contribution to the 
continuum lowering is described by the electronic SF $S_{\rm ee}^0$ and has been 
widely discussed, see Refs. \cite{KKER,KSK}. Results in the Montroll-Ward approximation are well known. 
% HR: ee-Term, Berechnung unklar dargestellt
% This includes the Debye shift $-\kappa e^2/(8 \pi \epsilon_0)$, for which an 
% additional factor $3/2-\pi/16$ arises from the exact evaluation 
% in the low-density limit, see Eq.\,(4.167) in Ref. \cite{KKER}. 
Compared to the ionic contribution $(\sim Z_i^2 e^2)$,
the electronic contribution $(\sim e^2)$ is usually quite small for highly charged states.

Following Eqs.~\eqref{deltasigmaeps} and~\eqref{Imeps},
we now discuss the ionic contribution to the correlation shift of the continuum edge 
$\repart \rmupper{\Sigma}{corr}{e}(0,\omega)+\repart \rmupper{\Sigma}{corr}{i}(0,\omega)$.
It is expressed as
\begin{eqnarray}
\label{deltaii}
&& \repart \rmupper{\Sigma}{corr,\, ion}{c}(p=0,\omega)= \rmupper{\Delta}{ion}{c}(0,\omega) 
\\&& = - {\cal P}\!\! \int \!\frac{d^3 {\bf q}}{(2\pi)^3}\!
 \int \!\frac{ d \omega'}{\pi} 
 \frac{V_{cc}(q)}{\omega-\omega'-E_{c,{\bf q}}/\hbar} 
  \frac{\pi Z_i e^2\, n_e}{\hbar \epsilon_0\, q^2} 
  \rmboth{S}{ZZ}{\rm ii}(q,\omega') \,.\nonumber
\end{eqnarray}
Thus, the ionic contribution to the continuum shift is related to the dynamical SF of the ions.
The quasiparticle shift has to be defined self-consistently at 
$\omega = \rmupper{\Delta}{ion}{c}(0,\omega)$, but this shift is 
compensated in the denominator of the integrand by the energy $E_{c,{\bf q}}$, which is shifted too. 
Then the ionic contribution $ \rmupper{\Delta}{ion}{c}(0,\omega)$ is given by $\rmupper{\Delta}{ion}{c}(0,0)$, 
later denoted as $\rmupper{\Delta}{ion}{c}$.

\subsection{Plasmon pole approximation}
Under WDM conditions considered here, the ions are strongly coupled, so that the SF $S_{\rm ii}^{ZZ}$ should 
not be taken in the Debye limit.
However, the plasma ions can be treated classically. Therefore, for $\rmupper{\Delta}{ion}{c}(0,0)$, see Eq.~\ref{deltaii}, 
we consider the limit $\hbar \rightarrow 0$ 
in the propagator $1/[-\omega'-\hbar q^2/(2 m_c)]$. In addition, the ions move very slowly in comparison to the electrons,
which indicates that it is reasonable to replace the dynamical SF of ions by the static SF within 
some approximations.
We use the plasmon pole approximation
${\rm Im}\, \varepsilon^{-1}_{\rm ion}(q, \omega)
=-\pi \omega_i^2 \{ \delta (\omega -\omega_{q,i})- \delta (\omega +\omega_{q,i})\}/(2\omega_{q,i})$,
where $\omega_{q,i}^2= \bks{q^2\,\omega_i^2}/\bks{\kappa_i^2\, \rmboth{S}{ZZ}{\rm ii}(q)}$ is fulfilling the f-sum rule~\cite{GRHGR07}
with the ionic plasmon frequency $\omega_i^2=Z_i^2n_i e^2/(\epsilon_0 m_i)$ 
and the inverse Debye screening parameter $ \kappa_i^2=\omega_i^2 m_i/k_BT$.
Then we find the following expression
\begin{eqnarray}\label{Szz}
&& \rmboth{S}{ZZ}{\rm ii}(q,\omega)\approx \rmboth{S}{ZZ}{\rm ii}(q)\,
\frac{ \delta (\omega -\omega_{q,i})+ \delta (\omega +\omega_{q,i})}{1+e^{-\hbar \omega /(k_BT)}}.
\end{eqnarray}
The physical meaning of the replacement of the dynamical SF by the static SF in Eq.~\eqref{Szz} is that 
the ions are considered to have a fixed distribution in the plasma neglecting temporal fluctuations.

For the ionization process $i_{_{Z_i}} \rightarrow e + i_{_{Z_i+1}} $, the IPD can be given by the
difference between the self-energy before and after the ionization of the investigated system, i.e.,
$ \rmboth{\Delta}{ion}{_{IPD}} = \rmupper{\Delta}{ion}{i}- (\rmupper{\Delta}{ion}{e}
+\rmupper{\Delta}{ion}{i+1})$.
We assume that the ionic structure of the plasma environment does not change during the ionization.
Therefore, we insert expression~\eqref{Szz} into $\rmupper{\Delta}{ion}{c}$, see Eq.~\eqref{deltaii}.
Performing the approximations as discussed in context with Eq.~\eqref{deltaii},
we obtain for the IPD 
\begin{equation}
\label{ionion}
 \rmboth{\Delta}{ion}{_{IPD}}=-\frac{(Z_i+1)e^2 }{2 \pi^2 \epsilon_0 } 
 \cdot \frac{\kappa_i^2}{k_{\mathrm{F},i}}
 \int_0^\infty\frac{d q_0}{q_0^2}\rmboth{S}{ZZ}{\rm ii}(q_0),
\end{equation}
where $q_0 = q/k_{\mathrm{F},i}$ is the reduced wavenumber with $k_{\mathrm{F},i} = \bks{3\pi^2 n_i}^{1/3}$.
Considering the ion-ion SF $ \rmboth{S}{DH}{ii}(q) =q^2/(q^2 + \kappa_i^2)$ of a
one-component plasma (OCP), valid  in the low density and the high temperature limits, the DH result 
$\rmboth{\Delta}{ion}{_{DH}} = - (Z_i+1)e^2\kappa_i/(4\pi\varepsilon_0)$ is recovered for the ionic contribution to the IPD.
The expression~\eqref{ionion} shows a strong dependence on the temperature indicated by the inverse Debye length $\kappa_i$ 
appearing in the frequency $\omega_{q,i} \sim \kappa_i^{-1}$ in Eq.~\eqref{Szz},
and also by the static ionic SF. The screening para\-meter $\kappa_i^{2} \propto 1/(k_BT)$
follows from the linearized Debye theory for classical systems.
Nevertheless, with increasing coupling parameter, the plasma starts to crystallize and forms a periodic structure.
In this case, the frequency $\omega_{q,i}$ is determined by
the Wigner-Seitz radius $r_{_{\rm WS}} =(4 \pi n_i/3)^{-1/3}$, as discussed, e.g., in Ref. \cite{Kalman}. Consequently, the parameter 
$\kappa_i^2$ occuring in $\omega_{q,i}$ should be replaced by a more general expression 
$\tilde \kappa_i^2(\Gamma_i)$ depending on the ionic coupling parameter $\Gamma_i=Z_i^2 e^2/(4 \pi \epsilon_0 k_BT r_{_{\rm WS}})$.
We can express  Eq.~\eqref{ionion} in the form
\begin{align}\label{final}
 \rmboth{\Delta}{ion}{_{IPD}}=-\frac{(Z_i+1)e^2}{2 \pi^2 \epsilon_0 r_{_{\rm WS}} }\cdot  S(\Gamma_i),
\end{align}
introducing the parameter function
\begin{equation} \label{sGamma}
 S(\Gamma_i) = F(\Gamma_i)
 \int_0^\infty\frac{d q_0}{q_0^2}\rmboth{S}{ZZ}{\rm ii}(q_0).
\end{equation}
From the Debye-H\"uckel theory follows
\begin{equation}\label{gerd}
 F(\Gamma_i) = \frac{\kappa_i^2\, r_{_{\rm WS}}}{k_{\mathrm{F},i}}= 
\bks{\frac{4}{9\pi}}^{1/3} r_{_{\rm WS}}^2\,\kappa_i^2 = \Gamma_i \bks{\frac{12}{\pi}}^{1/3}
\end{equation}
valid for weakly coupled system $\Gamma_i \ll 1$.
For strong coupling, a similar type of expression, $F(\Gamma_i) = \sqrt[3]{4/(9\pi)}\cdot r^2_{_{\rm WS}}\,  \tilde \kappa_i^2(\Gamma_i) $
can be defined and will be discussed in detail in the next section. One should keep in mind that, for a fixed charge state $Z_i$,
the parameter function $S(\Gamma_i)$ 
should gradually tend to a constant due to crystallization of the plasma with increasing coupling parameter~\cite{Kalman}.
At a fixed temperature and density, the parameter function $S(\Gamma_i)$ slightly depends on the charge number
since the dependence on charge number $Z_i$ in the static ionic SF $\rmboth{S}{ZZ}{\rm ii}(q_0)$ compensates with that of
the function $F(\Gamma_i)$.

The approach, presented in this work, shows a close connection of the IPD to the detailed structure of the plasma system.
The general expression \eqref{final} with \eqref{sGamma} should work within the valid range of the fluctuation-dissipation theorem
for both equilibrium and non-equilibrium systems described by the static SF of the quantum many-body system.
Once the SF is known from other methods, for instance, simulations or 
Thomson scattering measurements, the IPD can be directly evaluated.
In this work, the local thermodynamic equilibrium is assumed for the calculation. 
Further investigations are needed to describe non-equilibrium situations, for instance, 
after irradiations by strong short-pulse laser beams.

\section{Results and discussion}
\subsection{Model calculation: comparison to other approaches}
To determine the function $F(\Gamma_i)$ in the parameter function $S(\Gamma_i)$, Eq. (\ref{sGamma}), the implicit normalization relation~\cite{KKER}
\begin{equation}\label{nonLinear}
  \int_0^\infty\! dx \cdot\! x^2 
  \biggcurlybra{\! 1 \! - \! \exp\squarebra{-\frac{\Gamma_i}{x}\! \cdot \!\exp\bks{\!-\tilde \kappa_i(\Gamma_i)r_{_{\rm WS}}\,x  }\! }\! }
 \! =\! \frac{1}{3}
\end{equation}
according to the non-linear Debye theory is used, which avoids negative densities 
of the screening cloud.
The Debye-H\"uckel theory can be recovered by 
expanding the exponential function outside of the square brackets up to the first order in $\Gamma_i$,
see Eq.~\eqref{gerd}. 
For intermediate and strong coupling, 
Eq.~\eqref{nonLinear} has to be solved numerically.
In this work, we introduce the following expression
\begin{equation}\label{temDepen}
 F(\Gamma_i) = \sqrt[3]{\frac{4}{9\pi}} r^2_{_{\rm WS}}\,  \tilde \kappa_i^2(\Gamma_i) 
 = \frac{3 \Gamma_i}{ \sqrt{(9\pi/4)^{2/3} +3 \Gamma_i}}
\end{equation}
as an approximation which reproduces the Debye-H\"uckel limit~\eqref{gerd} as well as the numerical solutions of Eq.~\eqref{nonLinear} in the
strong coupling regime of interest.

In general, the pair correlation function exhibits a peak near  
$r_{_{\rm WS}}$ when approaching the liquid state, which would be reasonably well described 
by a Percus-Yevick SF. 
%\cite{PY58}. 
In the intermediate density region, an interpolation formula for the ionic SF can be applied, see Ref. \cite{GRHGR07}.
In the following, we use  expression \eqref{final} together with Eq.~\eqref{temDepen} and
the static ionic SF as given in Ref. \cite{GRHGR07} to 
evaluate the ionic contribution to the IPD in the plasma.

\begin{figure}[ht] 
\centering 
%\subfigure[Model calculation for the experiment \cite{Hoarty13}] { \label{fig:hoarty} 
\includegraphics[width=0.48\textwidth]{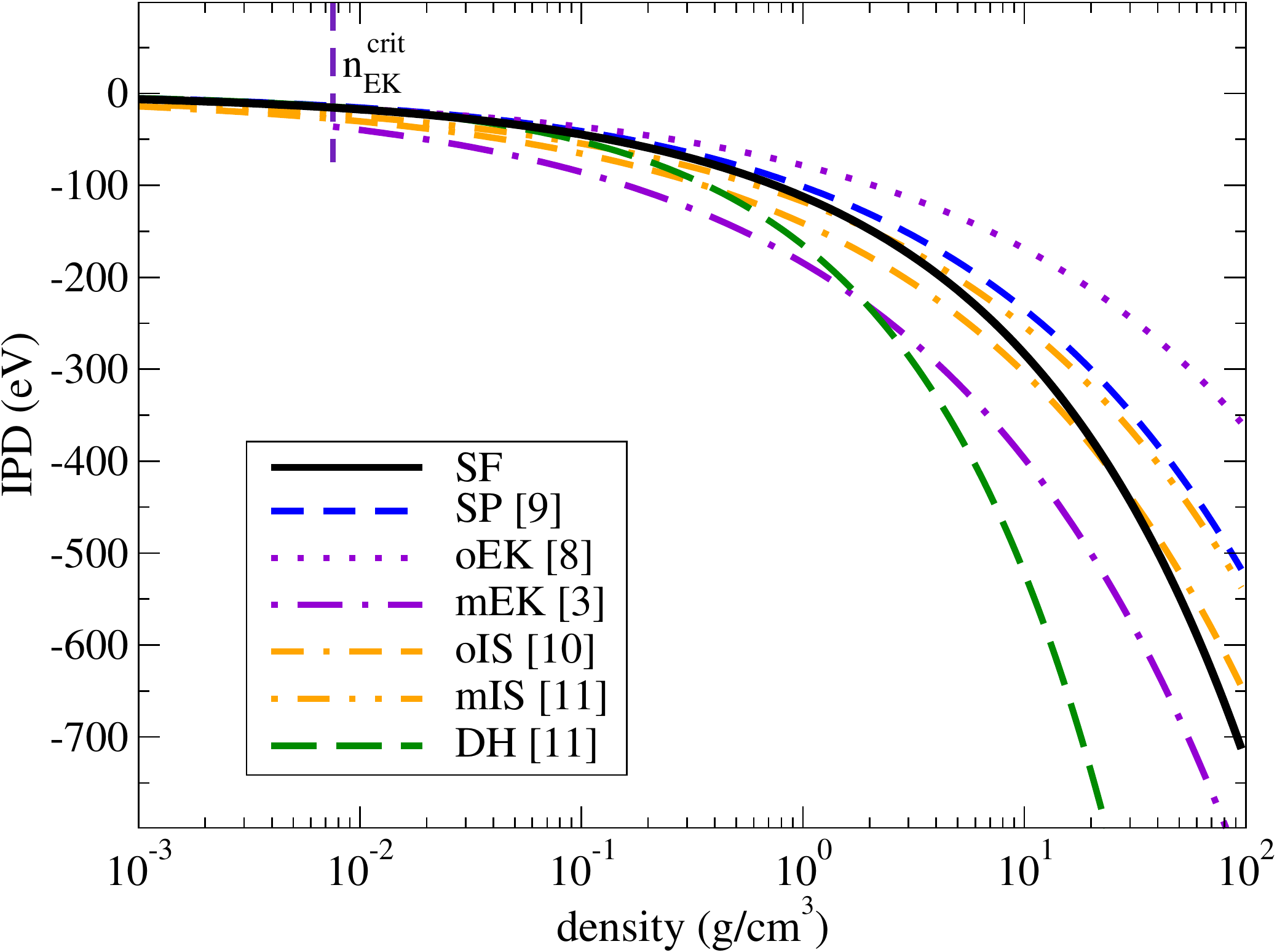}
 % \subfigure[Model calculation for the experiment \cite{ciricosta12}] { \label{fig:ciricosta} 
% \includegraphics[width=0.45\textwidth]{ciricosta.pdf}
\caption{(color online) IPD  for Al$^{11+}$ at 600 eV as function of the density, calculated using our model (SF) and by different theoretical models. } 
\label{fig:0} 
\end{figure}

As an exploratory  calculation in order to compare to
 other theoretical models,
%After that, detailed investigations for different experimental measurements and 
%comparisons of our results with other theoretical approaches and experimental predictions are shown.
we consider the IPD of the ion Al$^{11+}$  $(Z_{i}=11)$ at a temperature of 600 eV. Fig. \ref{fig:0} 
shows the IPD calculated using different theoretical models.
It can been seen, that the IPD from SP \cite{SP66}, original EK (oEK) \cite{EK63} and our result are 
in good agreement with
the DH shift in the low density region. 
Above the critical density $\rmboth{n}{crit}{_{EK}}= 3/(4\pi)\cdot(4\pi\varepsilon_0 k_BT/( Z^2 e^2))^3$ with 
the nuclear charge $Ze$,
the underestimation of the IPD by the SP model and the overestimation by the modified EK (mEK) model 
\cite{ciricosta12} can be seen
in comparison to the original IS (oIS) model.
Note that, with increasing density, corresponding to increasing coupling parameter $\Gamma_i$
($\Gamma_i = 0.16$ for the density $0.001$ g/cm$^3$ and $\Gamma_i = 7.28$ for the density $100$ g/cm$^3$),
our result shows, on one hand, a transition from SP at low densities (weakly and moderately coupled) to 
mEK at large densities (strongly coupled), and, on the other hand, a good agreement with the oIS model 
in the intermediate density region.

% %%%%%%%%%%%%%%%%%%%%%%%%%%%%%%%%%%%%%%%%%%%%%%%%%%%%%%%%%%%%%%%%%%%%%
% %%%%%%%%%%%%%%%%%%%%%%%%%%%%%%%%%%%%%%%%%%%%%%%%%%%%%%%%%%%%%%%%%%%%%%
\subsection{Numerical results for experimental conditions}

We now discuss the application of our model calculation to conditions  observed in experiments.
In the experiments of Hoarty \etal \,\cite{Hoarty13,Hoarty13a}, the spectral lines emitted from Al$^{11+}$ were observed.  
The investigated density range is 1.2 to 9 g/cm$^3$ 
at electron temperatures in the range of 550 to 700 eV. The disappearance of the ${\rm He}_\beta$ line would be due to the dissolution of n=3 levels. 
The assumption of local thermodynamic equilibrium
is believed to be valid for the high densities \cite{Hoarty13a}, which implies the ionic coupling parameter is estimated to be 
in the range of 2-4. In such a moderate coupling regime, the SP and IS models should result in the best
agreement with the experiment, as can be seen by looking at the relevant density range in  Fig.~\ref{fig:0}.

\begin{figure}[ht] 
\centering 
%\subfigure[Model calculation for the experiment \cite{Hoarty13}] { \label{fig:hoarty} 
\includegraphics[width=0.45\textwidth]{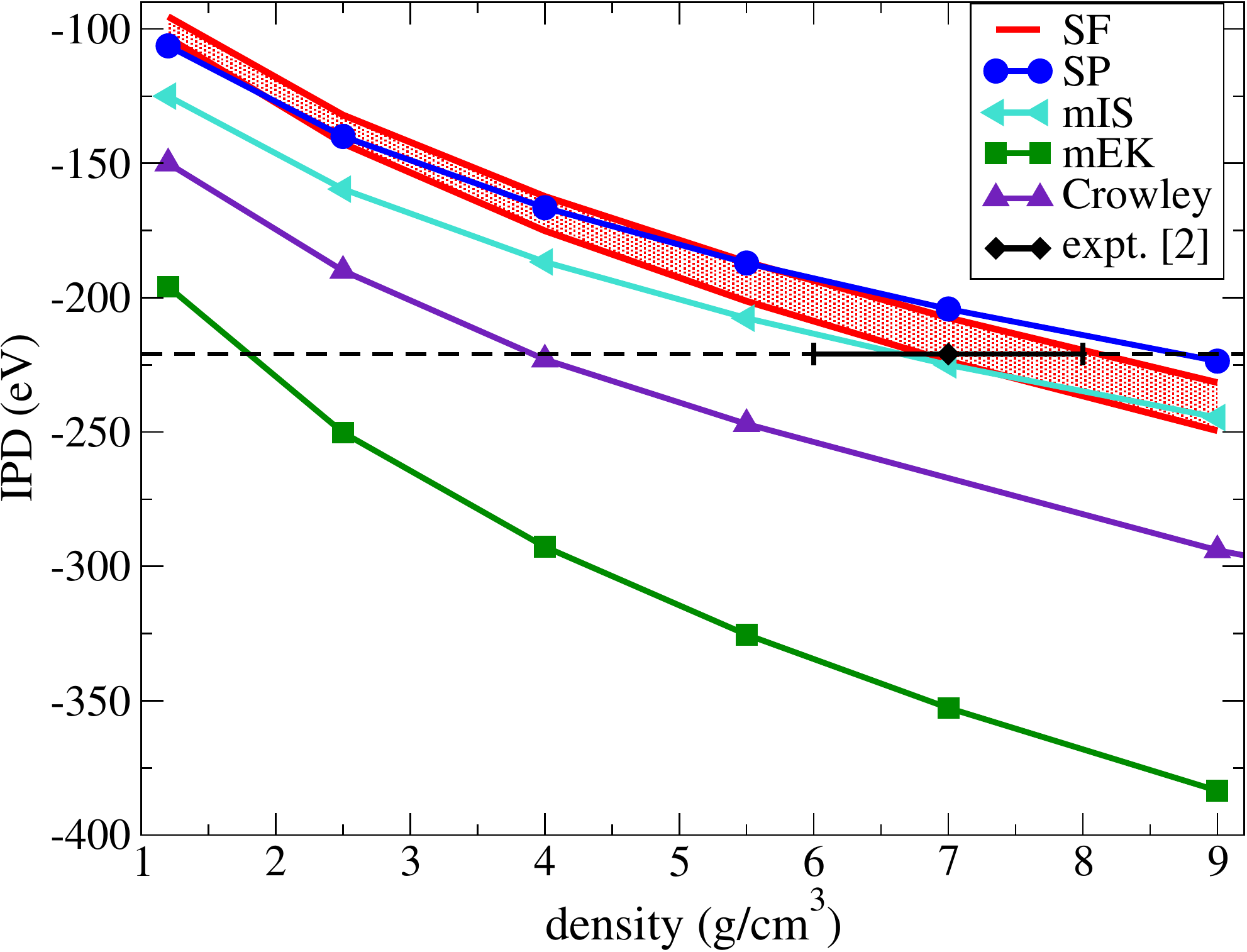}
 % \subfigure[Model calculation for the experiment \cite{ciricosta12}] { \label{fig:ciricosta} 
% \includegraphics[width=0.45\textwidth]{ciricosta.pdf}
\caption{(color online) IPD  for Al$^{11+}$ in aluminum plasma at 600 eV for the density as relevant for the experiment \cite{Hoarty13}, 
calculated by different theoretical models, i.e., SP~\cite{SP66}, mEK~\cite{ciricosta12}, 
and mIS as well as Crowley's calculation~\cite{Crowley14}.
The horizontal line indicates the unperturbed ionization
potential of the upper level of the Al$^{11+}$ ${\rm He}_\beta$ line~\cite{nist}. 
The full line (diamond with error bar) marks the critical density range observed experimentally \cite{Hoarty13}. 
The (red) shaded area shows the result from our model (SF) for temperatures in the range of 550 to 700 eV.}
%
%The experimental value in aluminum plasma is given where the Al$^{11+}$ ${\rm He}_\beta$ line fades out \cite{Hoarty13}.
%The corresponding results are different for different models. Our calculation for the IPD (the red region) shows the results
%for the electron
%temperature in the range of 550 to 700 eV. \CL{The black dashed line denotes the unperturbed ionization
%potential of the upper level of the Al$^{11+}$ ${\rm He}_\beta$ line~\cite{nist}. The results of Crowley are 
%taken from Ref.~\cite{Crowley14}.}} 
\label{fig:1} 
\end{figure}

The latter is shown in Fig.~\ref{fig:1} for a more detailed discussion.  The horizontal line denotes the unperturbed ionization
potential (220 eV) of the upper level of the Al$^{11+}$ ${\rm He}_\beta$ line~\cite{nist}. The density range,  
in which the disappearance of ${\rm Ly}_\beta$ and ${\rm He}_\beta$ lines in aluminum plasma \cite{Hoarty13} was 
measured, is marked as solid line.  It occurs at a density somewhere between 5.5 and 9 g/cm$^3$, which is in reasonable
agreement with the predictions by FLYCHK \cite{Chung05} using the SP model. 
According to calculations based on a generalized ion-cell
model by Crowley \cite{Crowley14}, for this range of densities, the modified IS (mIS) model is most suitable. 
This  is consistent with  predictions for spectra %for the opacity 
using the CASSANDRA opacity code with an IS model for the IPD \cite{Hoarty13a}, where the 
dissolution of lines from $n=3$ levels is indicated to take place between the density of $6\sim 8$ g/cm$^3$.
As shown in Fig.~\ref{fig:1}, the EK model
results in much larger IPD values  in comparison to the SP model,
and hence leads to a disappearance of spectral lines at a lower density of about 2 g/cm$^3$. A similar estimate
was given in the calculation by Crowley~\cite{Crowley14}.
Our approach, predicting a critical density between $7\sim 8$ g/cm$^3$ for the disappearance of $n=3$ levels, 
gives also an excellent agreement with the experimental data and with the predictions by the CASSANDRA opacity code~\cite{Hoarty13a}.

%The fact, that the agreement of our results changes gradually from the SP model to the mEK model with 
%increasing coupling parameter,
%gives a reasonably good explanation for the other experiment. 
Using 
FEL \cite{ciricosta12,ciricosta16}, further experiments have been performed recently. An aluminum 
sample at solid density was isochorically heated up to electron temperatures of
200 eV, indicating a strongly coupled plasma, and  the IPD was directly measured for different charge states.
The LCLS pulse duration in this experiment was estimated to be less than 80 fs. The ionic plasma frequency
$\rmboth{\omega}{ion}{pl} $ in this laser-produced plasma is found to be in the order of $10^{14}$/s. In contrast,
the response of the electrons to the laser field is much faster and can be described by the electron 
frequency $\rmboth{\omega}{el}{pl} \sim 10^{16}$/s. In comparison to the laser pulse, the electrons 
have enough time to exchange energy between each other and with the laser field and are isochorically heated to a high 
temperature. Because of the large mass of the ions, the response of the ionic subsystem to the external fluctuations 
is so slow that the ions in the plasma are weakly excited by the photons and by fast moving electrons, which implies that the ions  
are colder than the electrons. Of essential importance in the measurement is that the IPD for distinct charge states, 
inferred from the triggering energy of the photoionization, is measured at different time stages. This fact indicates 
that the ions are heated during the time evolution and  local thermal equilibrium  may be achieved.

\begin{figure}[ht] 
\centering 
% \subfigure[Model calculation for the experiment \cite{Hoarty13}] { \label{fig:hoarty} 
% \includegraphics[width=0.45\textwidth]{hoarty.pdf}
 %\subfigure[Model calculation for the experiment \cite{ciricosta12}] { \label{fig:ciricosta} 
\includegraphics[width=0.45\textwidth]{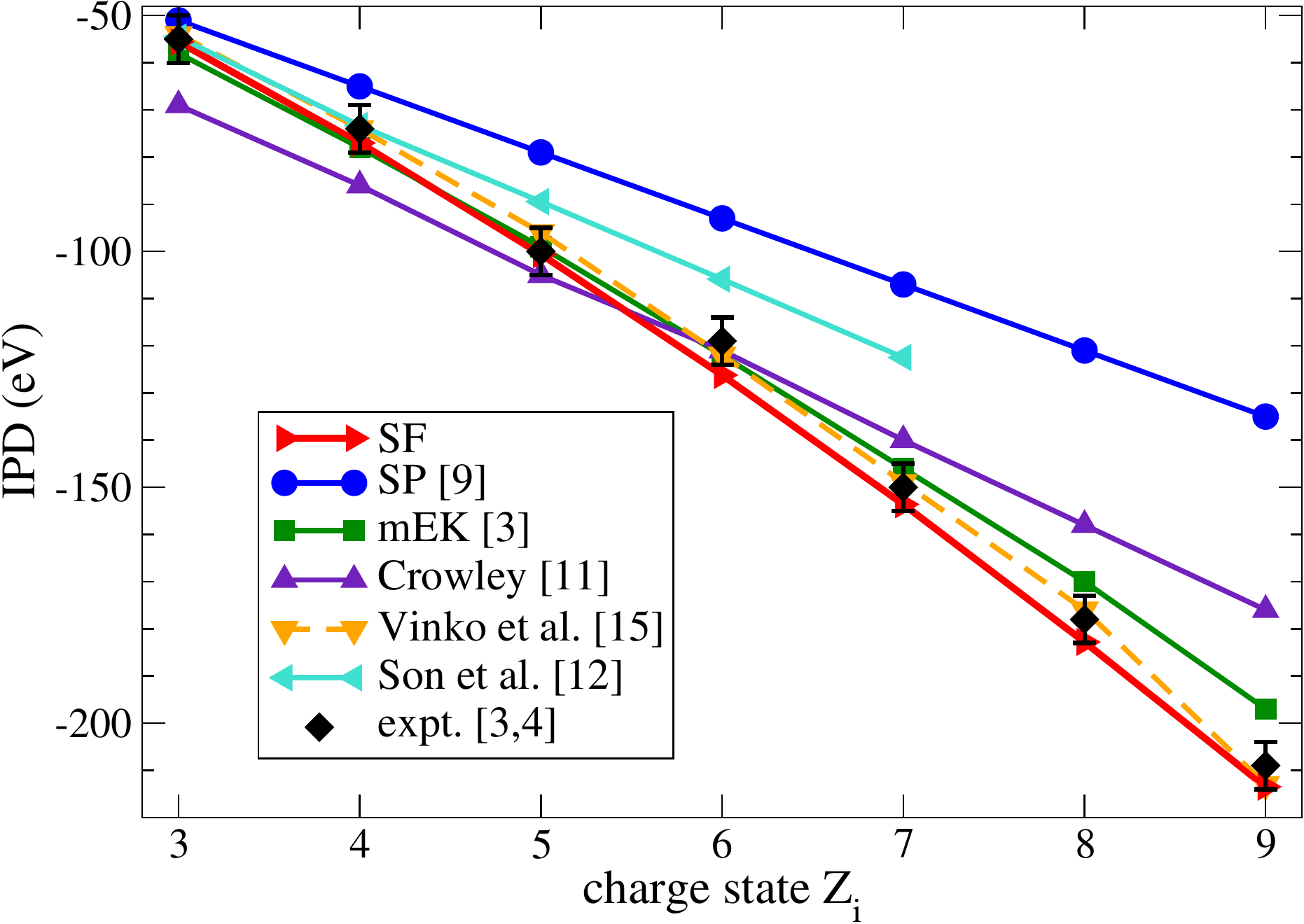} 
\caption{(color online) IPD  for aluminum plasma at solid density $2.7$ g/cm$^{3}$ as function of different charge states. 
Shown are experimental results  \cite{ciricosta12,ciricosta16} in comparison to  our model (SF) and  other theoretical models.
(Lines to guide the eye)} 
\label{fig:2} 
\end{figure}

Fig.~\ref{fig:2} shows the experimental results in comparison to several calculations using different theoretical models.
The direct measurement of the IPD in aluminum plasma  can be 
 explained reasonably well by the mEK model as discussed in Ref.~\cite{ciricosta12}. 
%This agreement is also shown in Fig.~\ref{fig:2}, where comparisons of results 
%evaluated from different theories are also given.
 Vinko \etal~\cite{VCW14} performed detailed calculations on
electronic structures of Al ions
in a plasma via the finite-temperature DFT method. They found that the IPD for a given charge state 
could be well
understood in
terms of the electronic structure of valence electron states near
core-excited ions within a pseudo-neutral atom approximation.
The results from the two-step Hartree-Fock calculations by Son \etal~\cite{STJZS14} and from the calculations by
Crowley~\cite{Crowley14} are less satisfying. However, as shown in Fig.~\ref{fig:2}, the experimental data can also be reproduced
by our approach, where the effect of the surrounding plasma on the ions is directly accounted for by the screened ionic SF.
In our calculation, the LTE condition was assumed. This might not be suitable for the experimental measurements where 
the ions remain relatively cold because of the femtosecond nature of the
X-ray pulse \cite{VCW14}. For this non-equilibrium case, the ionic SF under non-LTE conditions should be taken into account.
However, detailed calculations of the ionic SF in the non-LTE case are rather intricate and are still in progress.

% \begin{figure}[ht] 
% \centering 
% % \subfigure[Model calculation for the experiment \cite{Hoarty13}] { \label{fig:hoarty} 
% % \includegraphics[width=0.45\textwidth]{hoarty.pdf}
%  %\subfigure[Model calculation for the experiment \cite{ciricosta12}] { \label{fig:ciricosta} 
% \includegraphics[width=0.45\textwidth]{ciricostaTemp.eps} 
% \caption{Comparisons of the IPDs predicted by different models and the experimental results \cite{ciricosta12,ciricosta16}
% for an aluminum plasma at solid density $2.7$ g/cm$^{3}$. (Lines to guide the eye)} 
% \label{fig:2} 
% \end{figure}
% From this, it seems to be reasonable to conclude that the SP model is more appropriate for a weakly and 
% moderately coupled plasma, whereas
% the mEK model is more suitable for a strongly coupled plasma system. Actually, this statement is inconsistent with new 
% measurements on a CH mixture done at the NIF \cite{Kraus16}. The measured mean charge state can not be explained by either the SP 
% or the mEK models, as shown in Tab. \ref{kraus}. 
% Although the DH theory is inappropriate under the experimental conditions because of the strong
% coupling of the carbon ions $(\Gamma_i \sim 4)$, it results in larger IPDs and therefore gives a more 
% reasonable 
% agreement with the experiment than all other models, see Tab. \ref{kraus}.

The application of simple IPD models (e.g. SP) to a mixture of different ions is problematic as displayed by  recent 
measurements on a CH mixture at  NIF \cite{Kraus16}. The obtained mean charge state 
can not be explained by either the SP 
or the mEK models, as shown in Tab. \ref{kraus}.
Although the DH shift is inappropriate under the experimental conditions (strong
coupling of the carbon ions $(\rmnumb{\Gamma}{C} \sim 4)$), it results in larger IPDs and therefore gives a more reasonable 
agreement with the experiment than all other models.
This fact
can be attributed 
to the deficiency to account for
 strong correlation and fluctuation effects in these models. 
 
For the CH mixture, the influence of a different 
chemical species, the protons from the fully ionized hydrogen, on the properties of the carbon ions 
is, within  SP and EK models, described by an additional electron density. 
In our approach, this effect
can be more consistently taken into account by the ionic SF, which includes the response 
of all charged particles in the plasma.
We applied the linear mixing rule \cite{DG09} for the SF of a multi-component plasma, 
$\rmboth{S}{ZZ}{_{\rm ii}}(q_0) = x\, \rmboth{S}{ZZ}{_{\rm CC}}(q_0) + (1-x)\, \rmboth{S}{DH}{_{\rm HH}}(q_0)$. 
For the ratio  $x=0.75$  of carbon,  an estimated mean charge of 4.79 and therefore
%$(\mittel{\rmnumb{Z}{C}}{} = 4.82)$
 a close match with the experimental value of 4.92 $\pm$ 0.15 \cite{Kraus16} is found. 
 Under the experimental conditions \cite{Kraus16}, the carbon ions are strongly coupled while
the protons are weakly correlated.
The SF of the protons modifies 
the structure of the integrand in Eq.~\eqref{sGamma} leading to higher IPD values for the carbon ions,
and therefore push the carbon ions to a higher charge state.
\begin{table}[h] 
 \centering
 \begin{tabular}{ |c|c|c|c |c| }
  \hline
  \diagbox{model}{charge} &\phantom{a} ${\mathrm{C}^{3+}}$ \phantom{a}&\phantom{a} ${\mathrm{C}^{4+}}$ \phantom{a}
  &\phantom{a} ${\mathrm{C}^{5+}}$ \phantom{a}& mean charge \\ \hline
  DH & 261.3 & 326.7 & 392.0 & 4.91 \\
  SP & 91.7  & 108.3 & 123.9 & 4.18 \\
  IS & 103.2 & 119.7 & 135.2 & 4.21 \\
 mEK & 116.0 & 145.0 & 174.0 & 4.24 \\
  SF $(x=0.75)$ & 237.3 & 296.6 & 355.9 & 4.79 \\
  SF $(x=1)$  & 99.0 & 123.7 & 148.4 & 4.19 \\ \hline
expt. &      &    &     & 4.92 $\pm$ 0.15 \\ \hline
 \end{tabular}
\caption{IPD in eV and mean charge for CH mixture at density 6.74 g/cm$^{3}$ and $T = 86$ eV \cite{Kraus16}. 
The ionization energies for different charge
states are ${\rm I\, [C^{3+}] = 64.5\  eV, I\, [C^{4+}] = 392.1\  eV, I\, [C^{5+}] = 490.0\  eV}$. We have taken
in our calculation an effective SF, where $x$ is the carbon ratio.
%$\rmboth{S}{ZZ}{_{\rm ii}}(q_0) = x\, \rmboth{S}{ZZ}{_{\rm CC}}(q_0) + (1-x)\, \rmboth{S}{DH}{_{\rm HH}}(q_0)$, 
%the value $x=1$ and $x=0.75$, without and with the influence of protons, respectively. 
(For Refs. see Fig.~\ref{fig:0}.)}
\label{kraus}
\end{table}

Calculations for a pure C plasma at the same conditions (same ionic density of carbon and same temperature), 
lead to the mean ionization degree of 4.2.
For the CH plasma, the asymmetry of the charges and masses of  protons and carbon ions lead to strong fluctuations 
and hence significantly enhance the ionization.
Future discussions on experiments with pure C targets may test this effect.

More recently, a new experimental study on the ionization states of warm dense aluminum 
($T_e \sim 20 - 25$ eV and $\rho \sim 2.7$ g/cm$^3$) was performed~\cite{Mo17}. It was found that the observed time-dependent
absorption spectra are better described using the mEK model for the IPD than using SP and IS models.
This result agrees with our findings. For the given experimental conditions, the ion charge states Al$^{4+}$ and Al$^{5+}$ are
clearly seen, which indicates an ion coupling parameter
of $\Gamma_i \sim 7$. As discussed for Fig.~\ref{fig:0}, in such strongly coupled systems, the mEK model should lead to a better
description for the IPD.

%%%%%%%%%%%%%%%%%%%%%%%%%%%%%%%%%%%%%%%%%%%%%%%%%%%%%%%%%%%%%%
%%%%%%%%%%%%%%%%%%%%%%%%%%%%%%%%%%%%%%%%%%%%%%%%%%%%%%%%%%%%%
\section{Conclusions and further improvements}

We treated the in-medium two-particle problem (\ref{BSE}) within a quasiparticle approach
and obtained the contribution of the shift of the continuum edge to the IPD.
In addition to the continuum edge, also the bound state energy levels are shifted. 
Although their shifts are small as compared to the continuum lowering, see
\cite{Seidel,RKKKZ78,ZKKKR78,KKER,KSK}, these bound state level shifts should also be 
considered in a detailed calculation for the IPD. Note that the shift of bound state levels
has been observed in the shift of spectral lines, and quantum statistical calculations \cite{GHR91} 
agree well with experimental data.  

A more serious problem is the use of the quasiparticle approximation. Within a 
sophisticated Green function approach, the quasiparticle propagators are replaced by 
spectral functions, see, e.g., \cite{Fortmann09}, which describe also the finite 
life time of the quasiparticle excitations.
This leads to the fact that the energy gaps between the 
optical lines describing bound state transitions are washed out (Inglis-Teller 
effect \cite{ingtel39}).

In his monograph, Griem \cite{Griem64} described the broadening of spectral 
lines by the Stark effect leading to a shift of the observed series limit. The 
latter is described by 
\begin{equation}\label{slimit}
 n_s^{z-1}=\frac{1}{2}z^{3/5}(a_0^3N_e)^{-2/15}\,,
 \end{equation}
with $n_s$ - main quantum number, $N_e$ - electron number density. Eq.~(\ref{slimit}) 
was determined by a fit to a Holtsmark profile \cite{holts19} and corresponds 
to Eq.~(4) in Ref.~\cite{ingtel39}. Griem mentions that the shift of the series 
limit where lines fully overlap does not have a direct relation to the lowering of the ionization 
potential (last paragraph of section 5.7 in \cite{Griem64}). As discussed in Sec. \ref{sec:2}, definite values for the plasma parameters, where the ionization potential vanishes, can only be given 
within a quasiparticle (mean field) approximation which gives sharp energy levels.
As soon as the imaginary part of the effective Hamiltonian (\ref{Veff}) is taken into account,
the sharp energy levels become broadened as a consequence of their finite life time 
owing to collisions with the plasma particles.
Consequently, the rigorous discrimination between bound states (having a 
finite life time) 
and continuum states (including resonances)  
is no longer possible, and, strictly speaking, the concept of IPD based on sharp quasiparticle energy levels
 becomes obsolete.

We performed exploratory calculations using a simple model for the dynamical SF (\ref{Szz}).
As a main result, we found, that correlations which are described by the ionic SF are indeed  relevant for the IPD. 
As proposed, it would be of interest to perform experiments with pure substances like C. 
Compared to the large IPD seen in CH experiments \cite{Kraus16}, a lower IPD is expected for a pure C plasma. 
More details of the ionic subsystem 
may be incorporated, in particular the relaxation of the ionic subsystem and collective excitations 
(plasmons, phonons) can be treated. 
For a discussion see also Ref. \cite{Crowley14}.

Our approach is based on a Born approximation for the interaction of the two-particle system with the plasma ions.
The internal structure and dynamics of the plasma is described by the dielectric function which contains 
the polarization function $\Pi ({\bf q}, \omega)$,
\begin{equation}
\varepsilon({\bf q},\omega)=1-\frac{1}{\epsilon_0 q^2} \Pi ({\bf q}, \omega)
\end{equation}
Improving the RPA expression for the polarization function, two-particle correlations are included, see also \cite{RD}.
In particular, the ionic dynamical structure factor is taken into account if the cluster decomposition of the polarization
function is considered, here the two-ion distribution. Similar approaches have been used for the optical spectra \cite{GHR91} 
where also a cluster decomposition of the polarization function has been considered.

This discussion gives  a conception of how to improve our approach. 
The Born approximation has to be completed accounting for multiple interaction (so-called T matrix).
A more general diagram for the self-energy looks like 
\begin{minipage}[t]{0.46\textwidth}
\begin{figure}[H]
\centering 
\includegraphics[width=0.4\textwidth]{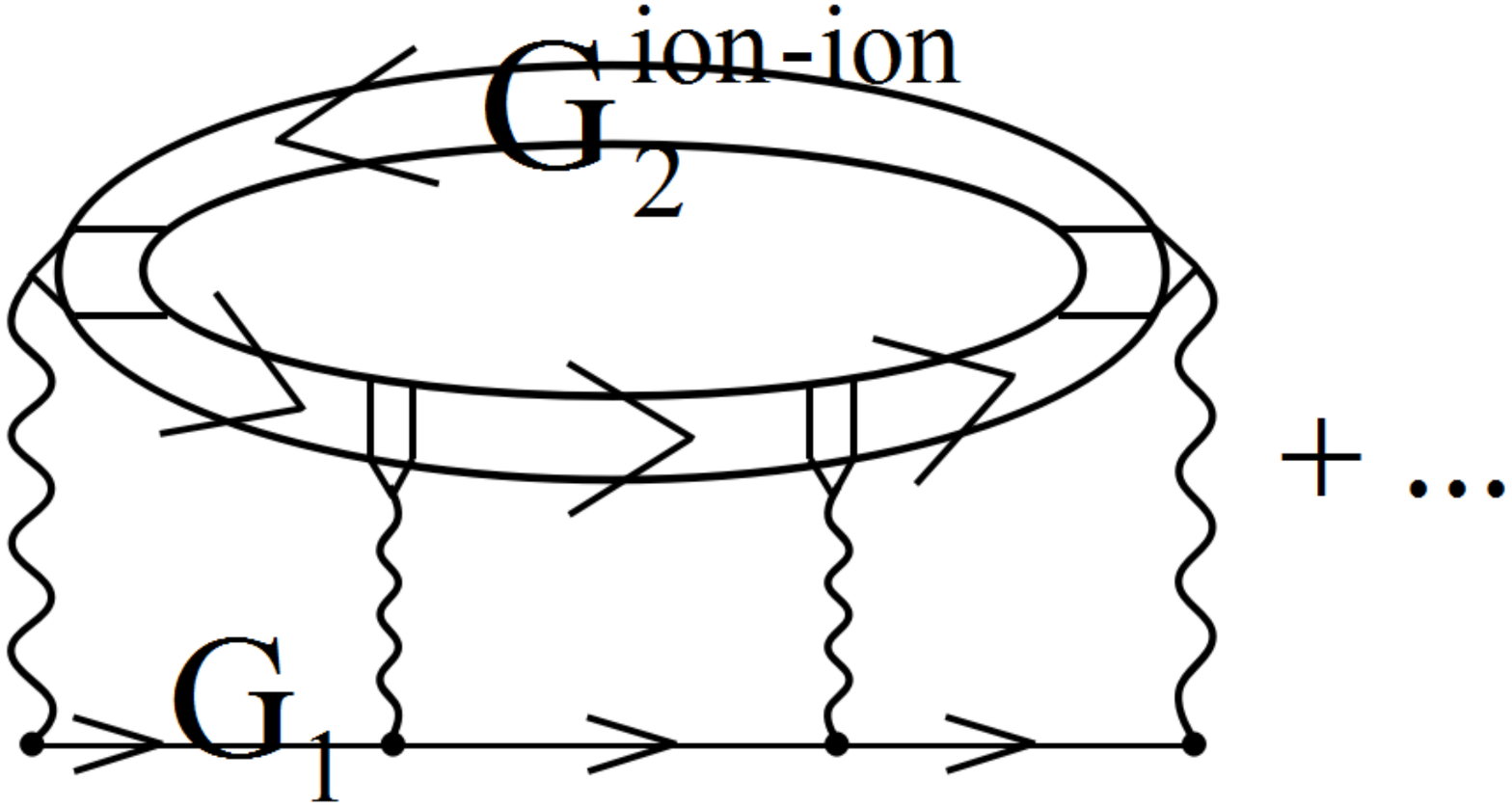}
\end{figure}
\end{minipage}
\noindent where the double line denotes the two-ion propagator, and the screened interaction 
with the investigated particle is considered in ladder approximation.
The approximation (\ref{GW}) for the self-energy results from the first contribution of 
the ladder sum which contains only two electron-ion interaction lines.

Starting from the general expression (\ref{GW}), we obtain a rather simple formula (\ref{final}) 
for the IPD containing the 
ionic static structure factor. 
We emphasize that this result could now be improved  by systematically removing again some of the approximations
for the dynamical SF (\ref{Szz}).
In particular, the plasmon pole approximation in handling the dynamical SF
is a model assumption which can be improved, e.g., by numerical simulations. 
Finally, an advantage of our quantum statistical approach is that any degeneracy effect can be taken 
into account in a systematic way, which becomes of interest at increasing densities.
%\section*{Acknowledgement}

{\it Acknowledgement:} One of the authors (CL) would like to thank S. Vinko for making experimental 
data available and D. Kraus, D. Hoarty, and Y. Hou for helpful discussions. 
This work is supported by the German Research Foundation DFG within SFB 652.

\end{document}